# Origin of magnetoresistance suppression in thin $\gamma$-MoTe$_2$

Shazhou Zhong[1,2], Archana Tiwari[1,2], George Nichols[1,2], Fangchu Chen[4,5], Xuan Luo[4], Yuping Sun[4,6,7], and Adam W. Tsen[1,3*]

[1]*Institute for Quantum Computing, University of Waterloo, Waterloo, Ontario N2L 3G1, Canada*
[2]*Department of Physics and Astronomy, University of Waterloo, Waterloo, Ontario N2L 3G1, Canada*
[3]*Department of Chemistry, University of Waterloo, Waterloo, Ontario N2L 3G1, Canada*
[4]*Key Laboratory of Materials Physics, Institute of Solid State Physics, Chinese Academy of Sciences, Hefei 230031, People's Republic of China*
[5]*University of Science and Technology of China, Hefei, 230026, China*
[6]*High Magnetic Field Laboratory, Chinese Academy of Sciences, Hefei 230031, People's Republic of China*
[7]*Collaborative Innovation Centre of Advanced Microstructures, Nanjing University, Nanjing 210093, People's Republic of China*

[*]Correspondence to: awtsen@uwaterloo.ca

**Abstract:**

We use both classical magnetotransport and quantum oscillation measurements to study the thickness evolution of the extremely large magnetoresistance (XMR) material and type-II Weyl semimetal candidate, $\gamma$-MoTe$_2$, protected from oxidation. We find that the magnetoresistance is systematically suppressed with reduced thickness. This occurs concomitantly with both a decrease in carrier mobility and increase in electron-hole imbalance. We model the two effects separately and conclude that the XMR effect is more sensitive to the former.

**Main text:**

Among the layered transition metal dichalcogenides (TMDCs), MoTe$_2$ is a unique member that crystallizes in both semiconducting 2$H$ and semimetallic 1$T$-type structures, making it an appealing candidate for novel phase-changing electronics [1]. It has already been demonstrated, for example, that transitions between the two can be controlled by strain, alloying, and electrostatic gating [2–6]. The semimetal polytype itself is interesting and exhibits different phases. First, true 1$T$ coordination is unstable as in-plane bond distortions dimerize the Mo atoms along the $b$-axis. Two stacking configurations of these distorted layers along the $c$-axis give rise to distinct three-dimensional (3D) structures: the centrosymmetric $\beta$ (or 1$T'$) phase at high temperature (above ~250K) and the noncentrosymmetric $\gamma$ (or $T_d$) phase at low temperature, with the difference being only a ~4° tilt in the unit cell. The latter structure notably hosts type-II Weyl nodes [7–13] and exhibits extremely large magnetoresistance (XMR) below ~20K [14].

XMR materials may be useful for spintronics and sensing applications. While both $\gamma$-MoTe$_2$ and $\gamma$-WTe$_2$, a structurally similar compound, have been shown to demonstrate XMR [14,15], its origin in the former is under debate. Transport studies have attributed the cause to a close compensation of electron and hole concentrations at low temperature for both materials [16–19]; however, angle-resolved photoemission experiments report that MoTe$_2$ remains uncompensated at all temperatures [20], in contrast to WTe$_2$ [21]. One can directly test the effect of charge (un)compensation on XMR in MoTe$_2$ by changing the relative carrier concentrations, but this is generally difficult to do in bulk systems without introducing unwanted disorder.

Recently, several of the authors have shown that the $\gamma$ phase is realized in thin MoTe$_2$

samples (below ~12nm) at all temperatures up to 400K [22], potentially allowing for the observation of Weyl nodes and their surface states under ambient conditions. The cause has been attributed to $c$-axis confinement of the hole bands. For the bulk crystal, it has been calculated that both electron and hole pockets shrink when cooling from the $\beta$ to $\gamma$ phase [23], and so it is possible that reducing thickness similarly stabilizes the latter at higher temperatures by confining the hole bands to lower energy.

In principle, these differences for thin samples should have a marked effect on the magnetoresistance (MR) at low temperature, provided that charge compensation is responsible for the XMR. Namely, we expect that a changing electronic structure would alter the delicate carrier balance achieved in the bulk crystal. We have performed both longitudinal and transverse magnetotransport measurements on MoTe$_2$ flakes at 300mK as a function of thickness. Not only do we observe lower MR in thin samples, fittings to a two-band model surprisingly show a decrease in both the absolute and relative carrier concentrations as well as their mobilities. This is qualitatively consistent with the measurement of Shubnikov-deHaas (SdH) oscillations, which provides an independent measure of the Fermi surface. By modeling the different effects separately, we conclude that the MR is more sensitive to changes in carrier mobility, and that, in principle, relatively large MR values can be achieved with a moderate degree of charge imbalance.

The upper inset of Fig. 1 shows an optical image of a representative device. Within a nitrogen-filled glovebox, we exfoliated MoTe$_2$ crystals onto polymer stamps and transferred the desired flakes onto oxidized silicon substrates with gold electrodes. The samples were subsequently covered with thin, insulating hexagonal boron nitride (hBN) before moving them outside the glovebox. This procedure prevents surface oxidation, which has been shown to affect the properties of other thin, metallic TMDCs [24–26]. In the main panel of Fig. 1, we show temperature dependent resistivity for three representative MoTe$_2$ samples of different thicknesses (7, 50, and 180nm) prepared in this manner. The traces are normalized to the resistivity at 280K and offset for clarity. All show metallic characteristics in contrast with an earlier report on unprotected thin flakes, which observes insulating behavior for thicknesses below ~10nm [14].

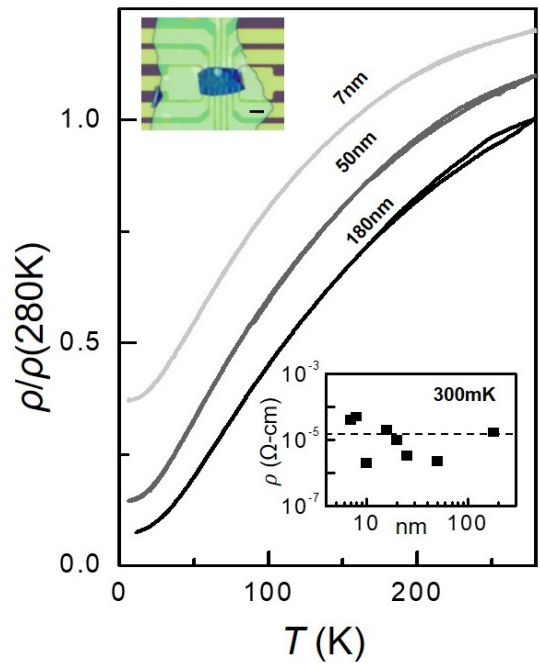

FIG. 1 Main panel: normalized resistivity as a function of temperature for three MoTe$_2$ flakes of different thicknesses. An offset was applied to the upper traces for clarity. Metallic behavior is observed down to 4K for all samples. Upper inset: optical image of a representative device. MoTe$_2$ is covered with thin hBN to prevent oxidation. Scale bar is 10μm. Lower inset: residual resistivity of all samples measured at 300mK. The dashed line marks the average resistivity.

As discussed in a previous work, the hysteresis loop associated with the $\beta - \gamma$ transition at ~250K gradually disappears for decreasing thickness as a single $\gamma$ phase is stabilized in thin flakes for the entire temperature range [22]. Below ~10K, the resistivity saturates to a temperature-insensitive, residual value. In the inset, we have explicitly plotted the residual resistivity for all of the samples measured in this work (eight in total) at 300mK. While there is variation between samples, we observe no direct trend with flake thickness. Furthermore, the average residual resistivity $1.8 \times 10^{-5} \Omega$-cm (marked by dashed line) is comparable to that of the bulk crystal (~$10^{-5} \Omega$-cm) [16], indicating that our flakes have not degraded during the preparation process.

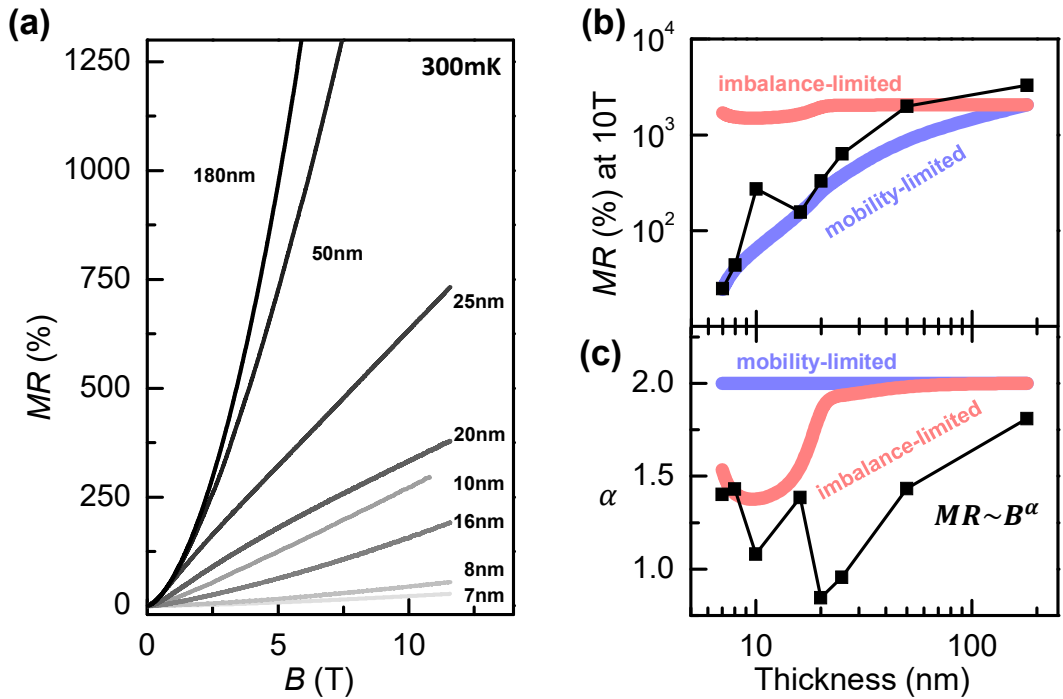

FIG. 2 (a) Percent magnetoresistance as a function of perpendicular magnetic field. MR decreases with decreasing thickness. (b) The MR measured at 10T for the different samples. (c) Exponent $\alpha$ of power law fit to the field dependence. Magnetoresistance is sub-quadratic as the thickness is reduced. The effects of charge imbalance and mobility decrease on MR and $\alpha$ are modeled separately and plotted in red and blue, respectively.

The MR, on the other hand, displays strong thickness dependent behavior. In Fig. 2(a), we plot $\mathrm{MR}(\%) = \frac{\rho(B)-\rho(0)}{\rho(0)} \times 100\%$, for the eight samples at 300mK and field applied along the $c$-axis. See Supplemental Material Fig. S1 for a discussion of symmetrization with respect to field direction. While the MR is always positive and unsaturating, we observe a clear and systematic suppression with reduced thickness. In the top panel of Fig. 2(b), we explicitly plot the thickness-dependent MR measured at 10T. The value is ~3000% in the 180nm flake, which is comparable to that measured in the bulk crystal [16], and decreases by two orders of magnitude in the thinnest samples.

Although the full Fermi surface of $\gamma$-MoTe$_2$ is complex and made up of multiple electron and hole pockets, the magnetotransport behavior may be understood from a simplified two-band model, where we assume that conduction takes place via one electron and one hole band only. Here, the field dependence of longitudinal and transverse resistivity obeys, respectively [15]:

$$\rho_{xx}(B) = \frac{(n\mu_n + p\mu_p) + (n\mu_p + p\mu_n)\mu_n\mu_p B^2}{e[(n\mu_n + p\mu_p)^2 + (p-n)^2 \mu_n^2 \mu_p^2 B^2]}, \qquad \rho_{yx}(B) = \frac{(p\mu_p^2 - n\mu_n^2)B + (p-n)\mu_p^2\mu_n^2 B^3}{e[(n\mu_n + p\mu_p)^2 + (p-n)^2 \mu_n^2 \mu_p^2 B^2]},$$

where $n$ ($p$) and $\mu_n$ ($\mu_p$) refer to the electron (hole) concentration and mobility, respectively. If $n = p$, the first equation simplifies to $\mathrm{MR} = \mu_n \mu_p B^2$, thus yielding unsaturating MR with a quadratic field dependence, as observed in bulk crystals [14,16,17]. The suppression of MR we observe in thin samples then suggests either (1) an imbalance of electrons and holes or (2) a decrease in their carrier mobilities. In order to distinguish between these two scenarios, we have first extracted the field dependent exponent from our data (MR~$B^\alpha$) using a power law scaling, and the result is plotted in the bottom panel of Fig. 2b. $\alpha$ is closer to 2 for thicker samples, but decreases with decreasing thickness, suggesting that the field dependent term in the denominator of $\rho_{xx}$ becomes more dominant, and thus $n \neq p$ for thinner flakes (scenario 1). The red and blue curves show the result of modeling the two scenarios separately, which will be discussed below.

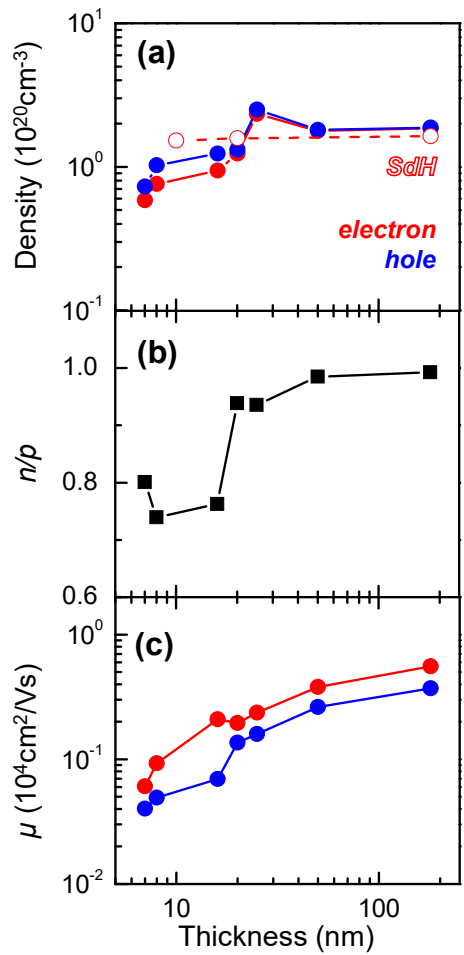

FIG. 3 (a) The carrier densities and (c) mobilities for different thickness samples extracted from the simultaneous fit of $\rho_{xx}$ and $\rho_{yx}$ using the two-band model. The electron-to-hole ratio for the different samples are shown in (b). As sample thickness decreases, concentration and mobility for both species decrease while additionally producing charge imbalance.

We would like to determine quantitatively the carrier concentrations and mobilities for the different thickness flakes. Unlike $\rho_{yx}$, however, $\rho_{xx}$ is insensitive to the carrier type. We have therefore measured both longitudinal and transverse resistivities in order to determine the full resistivity tensor of our samples, and the field (anti-)symmetrized results are shown in Supplemental Material Fig. S2 as solid lines. We have performed a simultaneous fit of $\rho_{xx}$ and $\rho_{yx}$ using the two-band model equations above (dashed lines in Fig S2), and the extracted carrier concentrations and mobilities are shown in Fig. 3, along with the relative carrier concentration $n/p$.

First, we note that the fit for $\rho_{xx}$ is consistently larger than the measured values at higher fields beyond 8T. The reason for this is that degree of carrier imbalance, $p - n$, allowed for by $\rho_{yx}$ is not sufficient to deviate $\rho_{xx}$ strongly away from a $B^2$ dependence. We have additionally used a three-band model to fit the data (see Supplemental Material Fig. S3), where we allow the charge of the third carrier to be a free parameter. While this indeed achieves a better fit, the third band has significantly lower mobility and fluctuates between electron and hole species. It is possible that it represents an effective average of the remaining bands, which acts to suppress further the MR at high fields. Nonetheless, we find that the behavior of the first two bands are qualitatively similar to that obtained in the two-band model, which suggests that this simplified picture may be sufficient to capture the physics.

Within the two-band model, we find that the density and mobility for both electrons and holes decrease slightly with reduced thickness. The ratio between electrons and holes,

$n/p$, also decreases, indicating greater carrier imbalance. For comparison, we have also plotted the electron densities determined by SdH oscillation measurements (open red circles), which will be discussed below. This behavior suggests that *c*-axis quantum confinement alone cannot account for the dimensionality-driven $\beta$ to $\gamma$ phase transition, as the electron pockets are mostly cylindrical [20,27]. It is consistent, however, with the transition energetics calculated by Kim *et al.* [23], which reports overall shrinking of both surfaces accompanying the $\beta$ to $\gamma$ transition. The mechanism behind the thickness-driven effect therefore remains open question.

Since both carrier imbalance and reduced mobility can suppress the MR, the two behaviors should be considered separately. In order to model this explicitly, we have taken the extracted carrier densities for the different thickness flakes and kept their mobilities constant and fixed to the values for the 180nm flake. We then calculated the MR within the two-band model using these new parameters, and plotted the MR percentage at 10T as well as field-dependent exponent on top of the original data in Fig. 2(b) in red ("imbalance-limited" curve). We similarly modeled the effect of reducing mobility only by fixing carrier densities fixed to the 180nm values. This is plotted in blue ("mobility-limited" curve). We observe that while subquadratic field dependence can be attributed to lack of charge compensation, the overall MR suppression is due to a reduction of carrier mobility. In principle, MR above 1000% may still be achieved in the thinnest samples, despite an imbalance ratio of $n/p\sim 0.8$, as long as the mobility can be made large.

At high fields, SdH oscillations have been observed in high-quality $MoTe_2$ crystals, which allows for an independent measurement of the Fermi surface [14,27,28]. We have observed oscillations for the 10nm, 20nm, and 180nm flakes at 300mK. Their background resistivities were subtracted (see Supplemental Material Fig. S4) and the results are plotted as a function of $1/B$ in the insets of Fig. 4(a). The samples show a beating pattern, thus indicating the presence of more than one frequency. We have taken the fast Fourier transform (FFT) of the SdH oscillations and the result is shown in the main panel of Fig. 4(a). We observe two clear peaks for all samples and possibly three for the 20nm flake, which show oscillations starting at lower field. For the 180nm flake, the peak positions (202T and 266T) are similar to what has been observed in the bulk crystal [27], and correspond to carrier densities of $0.70 \times 10^{20}$ $cm^{-3}$ and $0.93 \times 10^{20}$ $cm^{-3}$, respectively, via the Onsager relation. Density functional calculations indicate these oscillation frequencies are associated with electron pockets [27], while the hole pocket frequencies either exceed 1000T or fall below 100T. For the thinner flakes, these two peaks are systematically shifted to lower fields, and thus densities. In Fig. 3(a), we have plotted the sum of the densities extracted from the positions of the two prominent peaks (open red circles) in order to compare with the concentrations obtained from the two-band model. The thickness dependence is qualitatively consistent, although the electron density estimated from classical magnetotransport is lower in thinner samples. We have further performed the same measurement on the 20nm flake, which shows the most prominent oscillations, as a function of temperature (see Fig. 4(b)). By fitting the peak amplitudes to the Liftshitz-

Kosevich (LK) formula, we obtained an estimate of the effective masses (~1.0-1.2$m_0$), which are slightly larger than those measured for the bulk crystal (~0.8$m_0$) [27].

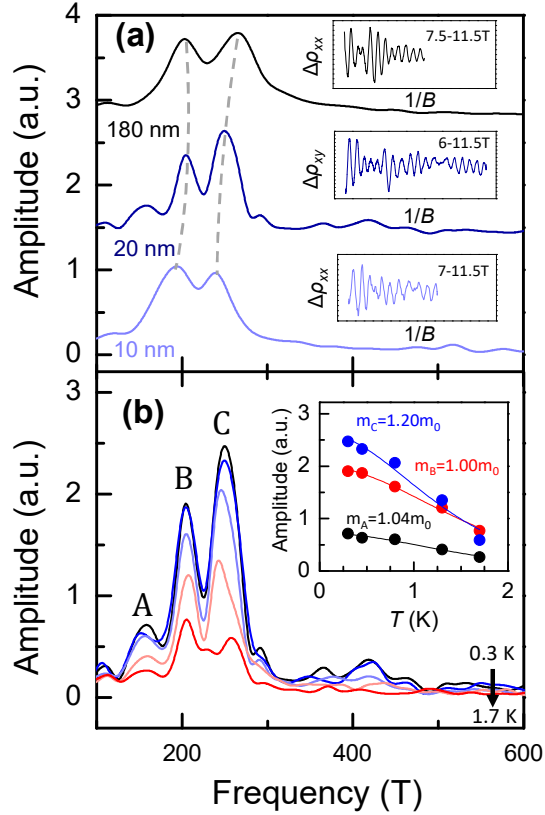

FIG. 4 (a) Inset: SdH oscillations observed after subtracting the polynomial background for each sample. Their corresponding FFT is shown in the main panel. The dashed lines guide the eye showing decreasing frequencies with decreasing sample thickness. (b) Main panel: The FFT of the SdH oscillations for the 20nm sample measured at 0.3K, 0.45K, 0.8K,1.3K, and1.7K. Inset: The amplitude of the FFT peaks as a function of temperature fit to the LK formula to extract the effective masses.

In summary, we used magnetotransport measurements to study the thickness evolution of electronic structure in $\gamma$-MoTe$_2$ at low temperature. We observe a decrease in both electron and hole densities as well as mobilities in thin flakes. The former possibly stabilizes the $\gamma$ phase in thin flakes at room temperature, while the latter greatly suppresses the XMR effect. At the same time, this has potentially interesting consequences for the tailoring of XMR in future materials. Namely, it may be possible to still achieve very large MR values without near-perfect charge compensation.

**Acknowledgements:**

We thank A. Burkov, R. Hill, and D. Rhodes for useful discussions. AWT acknowledges support from an NSERC Discovery grant (RGPIN-2017-03815). This research was undertaken thanks in part to funding from the Canada First Research Excellence Fund. Work in China was supported by the National Key Research and Development Program under contracts 2016YFA0300404 and the National Nature Science Foundation of China under contracts 11674326 and the Joint Funds of the National Natural Science Foundation of China and the Chinese Academy of Sciences' Large-Scale Scientific Facility under contracts U1432139 and Key Research Program of Frontier Sciences of CAS (QYZDB-SSW-SLH015).

SUPPLEMENTARY MATERIAL

# Origin of magnetoresistance suppression in thin $\gamma$-MoTe$_2$

Shazhou Zhong,[1,2] Archana Tiwari,[1,2] George Nichols,[1,2] Fangchu Chen,[4,5] Xuan Luo,[4] Yuping Sun,[4,6,7] and Adam W. Tsen[1,3*]

[1]*Institute for Quantum Computing, University of Waterloo, Waterloo, Ontario N2L 3G1, Canada*
[2]*Department of Physics and Astronomy, University of Waterloo, Waterloo, Ontario N2L 3G1, Canada*
[3]*Department of Chemistry, University of Waterloo, Waterloo, Ontario N2L 3G1, Canada*
[4]*Key Laboratory of Materials Physics, Institute of Solid State Physics, Chinese Academy of Sciences, Hefei 230031, People's Republic of China*
[5]*University of Science and Technology of China, Hefei, 230026, China*
[6]*High Magnetic Field Laboratory, Chinese Academy of Sciences, Hefei 230031, People's Republic of China*
[7]*Collaborative Innovation Centre of Advanced Microstructures, Nanjing University, Nanjing 210093, People's Republic of China*

[*]Correspondence to: awtsen@uwaterloo.ca

**Symmetrization of magnetotransport data**

The longitudinal resistivity, $\rho_{xx}$, and thus magnetoresistance, MR, should be symmetric with respect to magnetic field direction, while the transverse resistivity $\rho_{yx}$ should be antisymmetric. In practice, however, there is some deviation due to mixing between the two signals. We thus symmetrized $\rho_{xx}$ and anti-symmetrized $\rho_{yx}$ by: $\rho_{xx}(B>0) = \frac{\rho_{xx}(B)+\rho_{xx}(-B)}{2}$ and $\rho_{yx}(B>0) = \frac{\rho_{yx}(B)-\rho_{yx}(-B)}{2}$. Fig. S1 shows the original data for the 25nm sample, while the (anti-)symmetrized data are plotted as red dashed lines.

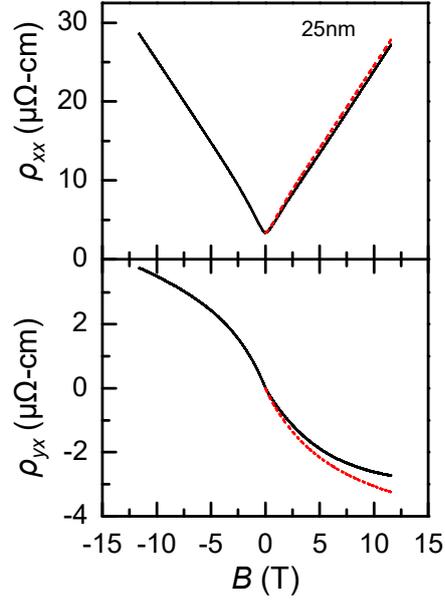

FIG. S1 The measured longitudinal and transverse resistivity for a representative (25nm) sample shown in the top and bottom panel, respectively. The (anti-)symmetrized data is plotted in red dashed lines.

**Two-band model fit**

The (anti-)symmetrized data are used to fit to the two band model:

$$\rho_{xx}(B) = \frac{(n\mu_n + p\mu_p) + (n\mu_p + p\mu_n)\mu_n\mu_p B^2}{e[(n\mu_n + p\mu_p)^2 + (p-n)^2 \mu_n^2 \mu_p^2 B^2]} \quad (1)$$

$$\rho_{yx}(B) = \frac{(p\mu_p^2 - n\mu_n^2)B + (p-n)\mu_p^2\mu_n^2 B^3}{e[(n\mu_n + p\mu_p)^2 + (p-n)^2 \mu_n^2 \mu_p^2 B^2]} \quad (2)$$

where $n$ ($p$) and $\mu_n$ ($\mu_p$) is the electron (hole) concentration and mobility, respectively. In Fig. S2(a) and (b), the solid lines show the experimental data and the dashed lines show the result of simultaneous fitting to equations (1) and (2). The extracted carrier densities, relative concentration ratio, and mobilities as a function of thickness are shown in Fig. S2(c).

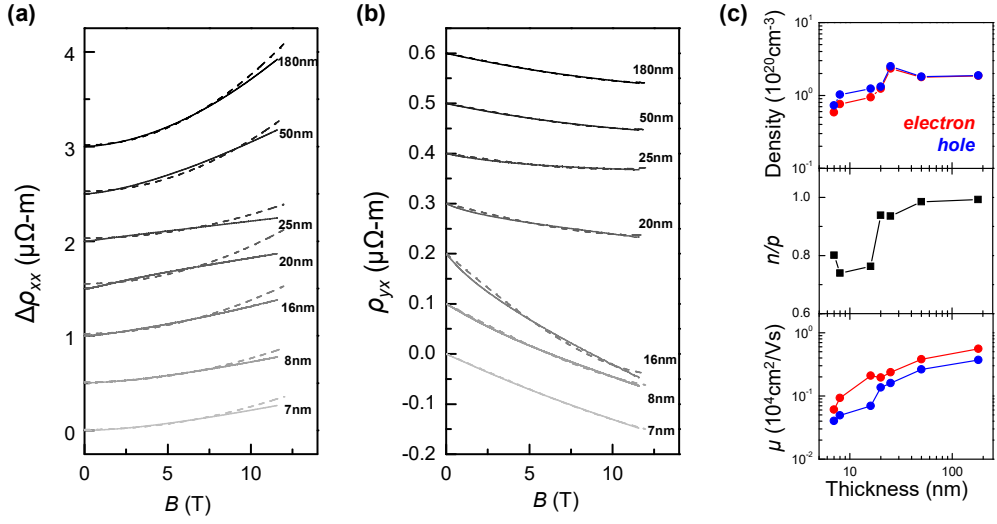

FIG. S2 (a) Symmetrized $\Delta\rho_{xx} = \rho_{xx}(B) - \rho_{xx}(0)$ and (b) anti-symmetrized $\rho_{yx}$ as a function of magnetic field for different sample thickness. Experimental data are shown as solid lines and two-band model fit are shown as dashed lines. Offset is applied to the traces for clarity. (c) Extracted carrier concentrations, electron-to-hole density ratio, and mobility as a function of thickness.

**Three-band model fit**

We have also fit the field-dependent resistivities using a three-band model. Here, it is more straightforward to consider the conductivities:

$$\sigma_{xx}(B) = \frac{ne\ \mu_n}{1+(\mu_n B)^2} + \frac{pe\mu_p}{1+(\mu_p B)^2} + \frac{n_3 e\mu_3}{1+(\mu_3 B)^2}, \quad (3)$$

$$\sigma_{xy}(B) = \left[-\frac{ne\mu_n}{1+(\mu_n B)^2} + \frac{pe\mu_p}{1+(\mu_p B)^2} \pm \frac{n_3 e\mu_3}{1+(\mu_3 B)^2}\right] eB. \quad (4)$$

The charge for third species $n_3$ is left as a free parameter. The above formulas can be inverted to obtain the resistivities:

$$\rho_{xx} = \frac{\sigma_{xx}}{\sigma_{xx}^2 + \sigma_{yx}^2} \quad (5)$$

$$\rho_{yx} = \frac{\sigma_{xy}}{\sigma_{xx}^2 + \sigma_{yx}^2}, \quad (6)$$

The results after simultaneous fitting of the experimental data to equations Eq. (5) and (6) are shown in Fig. S3.

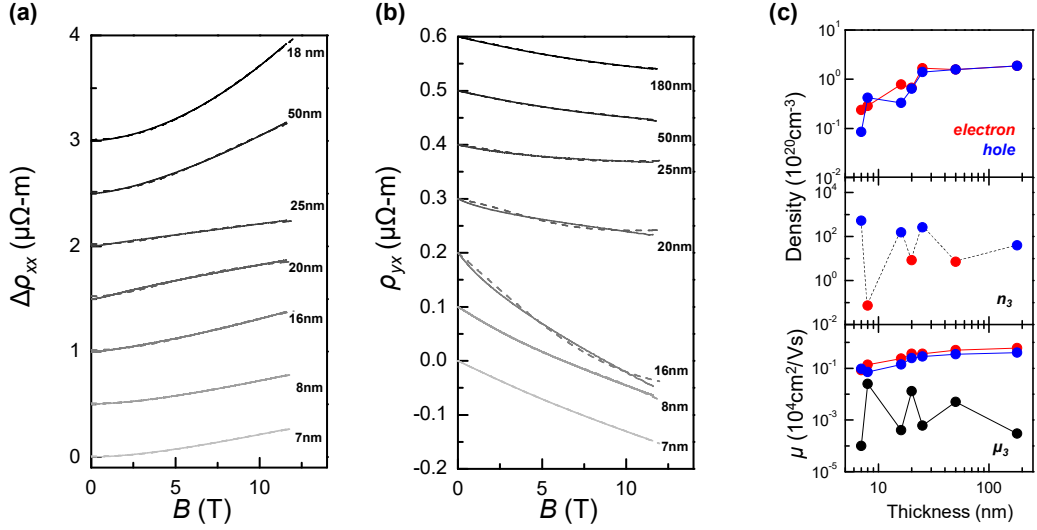

FIG. S3 (a) Symmetrized $\Delta\rho_{xx} = \rho_{xx}(B) - \rho_{xx}(0)$ and (b) anti-symmetrized $\rho_{yx}$ as a function of magnetic field for different sample thickness. Experimental data are shown as solid lines and three-band model fit are shown as dashed lines. Offset is applied to the traces for clarity. (c) Extracted carrier concentrations and mobility as a function of thickness.

**Quantum oscillations**

We observe quantum oscillations in $\rho_{xx}(B)$ or $\rho_{yx}(B)$ for three samples (10, 20, and 180nm). The upper panel of Fig. S4 shows representative data for the 20nm sample, while the lower shows the result after subtracting a polynomial fit (dashed red line).

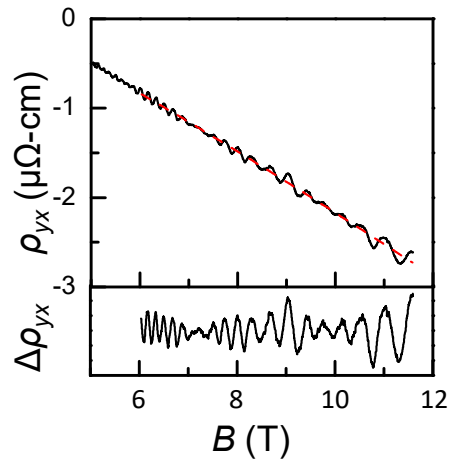

FIG. S4 (Top panel) Original $\rho_{yx}(B)$ data measured for the 20nm sample. (Bottom) result after subtracting a polynomial fit (red dashed line).